\documentclass[seceq]{ptptex}

\usepackage{graphicx}

\title{
{\normalsize\vspace*{-1.9cm}\\\hfill\mbox{ITP-Budapest 607}\\
\hfill\mbox{WUB 04-03}\\\vspace*{0.6cm}}
Finite $T/\mu$ lattice QCD and the critical point%
}

\author{
Z, \textsc{Fodor}$^{a,b}$, S.D. \textsc{Katz}$^a$\thanks{On leave from 
Institute for Theoretical Physics, E\"otv\"os University,
Budapest, Hungary.} %
}

\inst{
$^a$Department of Physics, University of Wuppertal, Germany\\
$^b$Institute for Theoretical Physics, E\"otv\"os University, 
P\'azm\'any 1, H-1117 Budapest, Hungary
}

\abst{
We present results on lattice QCD at finite temperature ($T$) and 
chemical potential ($\mu$). We apply the overlap-improving multi-parameter
reweighting technique for $n_f=4$ and $2+1$ staggered QCD. We use 
semi-realistic masses on $L_t$=4 lattices. The critical endpoint
(E) of 2+1 flavour QCD on the $\mu$-T plane is located. 
Our results are based on ${\cal{O}}(10^3-10^4)$ configurations.
}

\begin{document}

\maketitle

\section{Introduction}

QCD at finite $T$ and/or $\mu$ describes relevant features of 
particle physics
in the early universe, in neutron stars and in heavy ion collisions.
Extensive experimental work has been done
with heavy ion collisions at CERN and Brookhaven to explore
the $\mu$-$T$ phase boundary (cf. \cite{Braun-Munzinger:2001mh}). It is
a long-standing question, whether a critical point
exists on the $\mu$-$T$ plane,
and particularly how to tell its location theoretically
\cite{Halasz:1998qr}.

\begin{figure}[htb]
\centerline{\includegraphics[width=10cm]{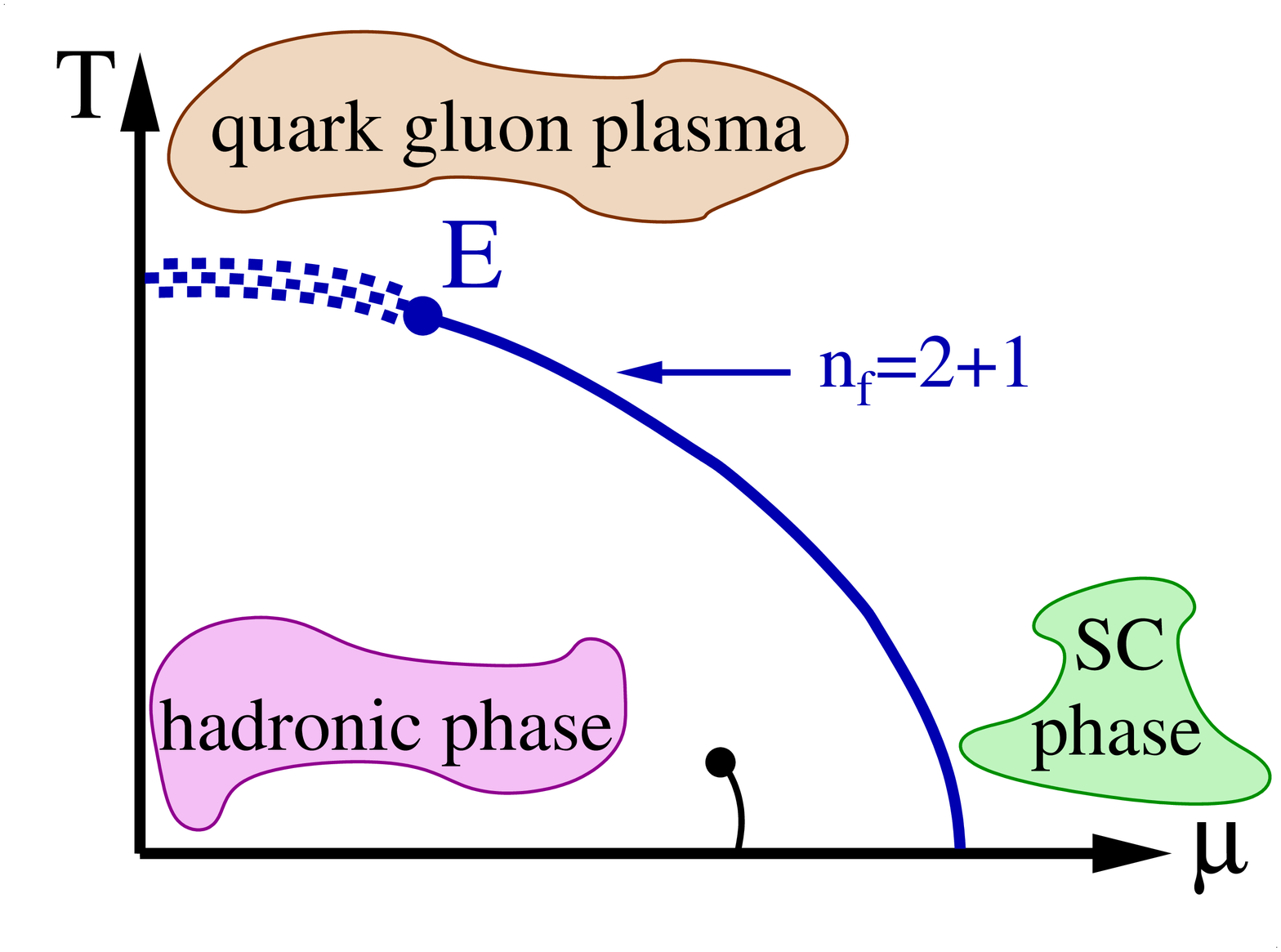}}
\caption[]{Schematic phase diagram of QCD in the $\mu-T$ plane. At $\mu=0$ 
and finite $T$ a cross-over is expected. The endpoint (E) connects the
cross-over region with the first-order region at large $\mu$. 
}\label{fig0}
\end{figure}

Let us start with the $\mu$=0 case first (see Fig.~\ref{fig0}).
Universality arguments \cite{Pisarski:ms} and lattice 
simulations \cite{Fodor:2002sd}
indicate that in a hypothetical QCD
with a strange (s) quark mass ($m_s$) as small as the up (u) and down (d)
quark masses ($m_{u,d}$)
there would be a first order finite
$T$ phase transition. The other extreme case ($n_f$=2)
with light u/d quarks but with an infinitely large $m_s$
there would be no phase transition, only a
crossover. Observables change rapidly during a crossover,
but no singularities appear.
Between the two extremes there is a
critical strange mass ($m_s^c$) at which one has a second order finite
$T$ phase transition. Staggered lattice results on $L_t$=4 lattices
with two light quarks and $m_s$ around the transition $T$ ($n_f$=2+1)
indicated \cite{Brown:1990ev} that $m_s^c$ is about half of the physical $m_s$.
Thus, in the real world we probably have a crossover.

At non-vanishing $\mu$, one realizes that arguments
based on a variety of models (see e.g. \cite{Barducci:1989wi,
Alford:1997zt,Halasz:1998qr})
predict a first order finite $T$ phase transition at large $\mu$.
Combining the $\mu=0$ and large $\mu$ informations an interesting
picture emerges on the $\mu$-$T$ plane. 
For the physical $m_s$
the first order phase transitions at large $\mu$ should be connected
with the crossover on the $\mu=0$ axis. This suggests
that the phase diagram features a critical endpoint $E$ (with
chemical potential $\mu_E$ and temperature $T_E$), at which
the line of first order phase transitions ($\mu>\mu_E$ and $T<T_E$)
ends \cite{Halasz:1998qr}. At this point the phase transition is of
second order and long wavelength fluctuations appear, which
results in (see e.g. \cite{Borsanyi:2001rc}) consequences, similar to
critical opalescence. Passing close enough to ($\mu_E$,$T_E$)
one expects simultaneous appearance of
signatures
which exhibit nonmonotonic dependence on the
control parameters \cite{Stephanov:1999zu},
since one can miss the critical point on either sides.

The location of E is
an unambiguous, non-perturbative prediction of QCD.
No {\it ab initio}, lattice analysis based on QCD was done to locate
the endpoint.  Crude models
with $m_s=\infty$ were used (e.g. \cite{Halasz:1998qr})
suggesting that $\mu_E \approx$ 700~MeV, which should be smaller
for finite $m_s$. The goal of our
work is to propose a new method to study lattice QCD at 
finite $\mu$ and apply it to locate the endpoint.
We use full QCD with dynamical $n_f$=2+1 staggered quarks.

QCD at finite $\mu$ can be given
on the lattice \cite{Hasenfratz:1983ba}; however, standard
Monte-Carlo techniques fail. At Re($\mu$)$\neq$0 
the determinant of the Euclidean Dirac operator is complex, which
spoils any importance sampling method.

Several suggestions were studied in detail to solve the problem.
We list a few of them (for a recent review see Ref. \cite{Philipsen:2001ws}).

In the large gauge coupling limit
a monomer-dimer algorithm was used \cite{Karsch:1988zx}.
For small gauge coupling an attractive
approach is the ``Glasgow method'' \cite{Barbour:1997ej} in which the
partition function is expanded in powers of $\exp(\mu/T)$
by using an ensemble of configurations weighted by the $\mu$=0 action.
After collecting more than 20 million configurations only unphysical
results were obtained: a premature onset transition.
The reason is that the $\mu$=0 ensemble does not overlap sufficiently
with the states of interest.
Another possibility is to separate the absolute value and the
phase of the fermionic determinant and use the former to generate
configurations and the latter in observables \cite{Toussaint:1989fn}. The curvature
of the $\mu-T$ phase diagram at $\mu=0$ was determined by a stochastic 
calculation of the derivatives of the fermion determinant 
in~\cite{Allton:2002zi}.

At imaginary $\mu$ the measure remains positive and standard Monte-Carlo
techniques apply. The canonical partition function can be obtained by
a Fourier transform \cite{imag,Alford:1998sd}. In this technique the dominant
source of errors is the Fourier transform rather than the poor overlap.
One can, however, use the fact that the transition line is an
analytic function of $\mu$, and a polynomial fit
for imaginary $\mu$ values could be analytically continued to real
values of $\mu$. The curvature of the phase diagram has been determined using 
this technique for 2,3 and 4 flavour staggered 
QCD~\cite{deForcrand:2002ci,deForcrand:2003hx,D'Elia:2002gd}.
At temperatures sufficiently above the transition,
both real and imaginary $\mu$ can be studied by
dimensionally reducing QCD \cite{Hart:2000ha}.
Hamiltonian formulation may also help studying the problem
\cite{Gregory:1999pm}.

\begin{figure}[htb]
\centerline{\includegraphics[width=10cm]{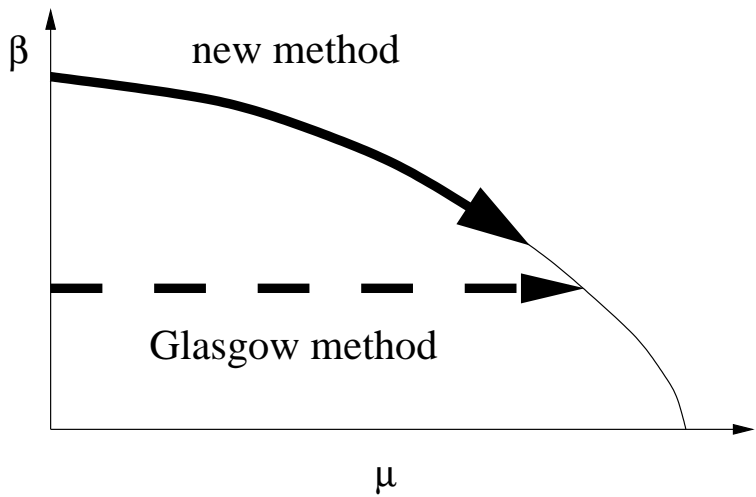}}
\caption[]{Schematic difference between the 
present and the Glasgow methods.}
\label{fig1}
\end{figure}

We propose a method
to reduce the overlap problem and determine the
phase diagram in the $\mu$-T plane (for details see \cite{Fodor:2001au}).
The idea is to produce an ensemble of QCD configurations at
$\mu$=0 and at the transition temperature $T_c$. Then we determine
the Boltzmann weights \cite{Ferrenberg:1989ui} of these configurations at $\mu\neq 0$
and at $T$ lowered to the transition temperatures at this
non-vanishing $\mu$. Since transition configurations
are reweighted to transition ones a much better
overlap can be observed than by reweighting pure had\-ronic configurations
to transition ones \cite{Barbour:1997ej}. Since the original 
ensemble is collected at $\mu$=0 we do not expect to be able to
decsribe the physics of the large $\mu$ region with e.g. exotic
colour superconductivity. Fortunately, the typical $\mu$ values
at present heavy ion accelerators are smaller than the region
we cover.

We apply our technique to $2+1$ flavour staggered QCD and
locate the critical point of QCD using semi-physical quark masses.
(Multi-dimensional reweighting
was successful for determining
the endpoint of the hot electroweak plasma \cite{Aoki:1999fi}
e.g. on 4D lattices.)

\section{Overlap improving multi-parameter reweighting}

Let us study a generic system of fermions $\psi$ and bosons $\phi$,
where the fermion Lagrange density is ${\bar \psi}M(\phi)\psi$.
Integrating over the Grassmann fields we get:
\begin{equation}\label{path_int}
Z(\alpha)=\int{\cal D}\phi \exp[-S_{bos}(\alpha,\phi)]\det M(\phi,\alpha),
\end{equation}
where $\alpha$ denotes a set of parameters of
the Lagrangian. In the case of staggered QCD $\alpha$
consists of $\beta$,
$m_q$ and $\mu$.
For some choice of the
parameters $\alpha$=$\alpha_0$
importance sampling can be done (e.g. for Re($\mu$)=0).
Rewriting eq. (\ref{path_int})
\begin{eqnarray}\label{reweight}
Z(\alpha)=
\int {\cal D}\phi \exp[-S_{bos}(\alpha_0,\phi)]\det M(\phi,\alpha_0)&& 
\nonumber \\
\left\{\exp[-S_{bos}(\alpha,\phi)+S_{bos}(\alpha_0,\phi)]
{\det M(\phi,\alpha)  \over \det M(\phi,\alpha_0)}\right\}.&&
\end{eqnarray}
We treat the curly bracket as an observable
(measured on each configuration)
and the rest as the measure. Changing
only one parameter of the ensemble
generated at $\alpha_0$ provides an accurate value for some observables
only for high statistics. This is ensured by
rare fluctuations as the mismatched measure occasionally sampled the
regions where the integrand is large. This is the
overlap problem. Having several parameters
the set $\alpha_0$ can be adjusted to get
a better overlap than obtained by varying only one parameter.


The basic idea of the method as applied to dynamical QCD can be
summarized as follows. We study the system at ${\rm Re}(\mu)$=0 around
its transition point. Using a Glasgow-type technique we calculate the
determinants for each configuration for a set of $\mu$, which, similarly
to the Ferrenberg-Swendsen method \cite{Ferrenberg:1989ui}, can be used for
reweighting.  The average plaquette values can be used to perform an
additional reweighting in $\beta$.  Since transition configurations were
reweighted to transition ones a much better overlap can be
observed than by reweighting pure hadronic configurations to transition
ones as done by the Glasgow-type techniques. The differences between the two
methods are shown in Figure \ref{fig1}. Moving along the transition
line was also suggested by Ref. \cite{Alford:1998sd}.

\begin{figure}[htb]
\centerline{\includegraphics[width=10cm]{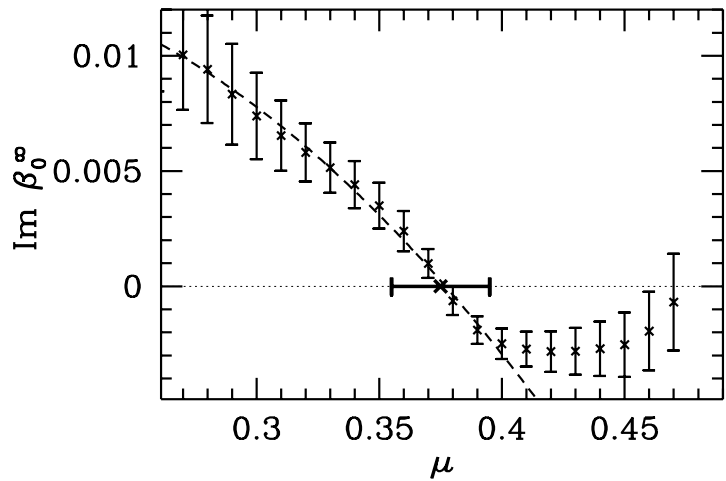}}
\caption[]{Im($\beta_0^\infty$) as a function of the chemical
potential.}
\label{fig3}
\end{figure}


These ideas have been tested using
imaginary chemical potential and a perfect agreement was found between 
direct simulations and multiparameter reweighting~\cite{eos_long}.
Based on these experiences we 
expect that our method can be applied at Re($\mu$)$\neq$0.

\section{The endpoint of $n_f=2+1$ QCD}

In QCD with $n_f$ staggered quarks
one should change the determinants to their $n_f$/4 power in our two
equations. Importance sampling works also in this case  at some $\beta$ and
at Re($\mu$)=0. Since $\det M$ is complex
an additional problem arises, one should
choose among the possible Riemann-sheets of the fractional power
in eq. (\ref{reweight}). This can be done by using \cite{Fodor:2001au}
the fact that at $\mu$=$\mu_w$ the ratio of the determinants is 1 and
it should be a continuous function of $\mu$.

In the
following we keep $\mu$ real and look for the zeros of $Z$
for complex $\beta$.  At a first order phase transition the free
energy $\propto \log Z(\beta)$ is non-analytic.
A phase transition appears only in the V$\rightarrow \infty$ limit,
but not in a finite $V$. Nevertheless, $Z$
has zeros at finite V, generating the non-analyticity of the
free energy, the Lee-Yang zeros \cite{Yang:be}.
These are at complex parameters (e.g. $\beta$). For a
system with first order transition these zeros
approach the real axis as V$\rightarrow \infty$ 
by a $1/V$ scaling.
This V$\rightarrow \infty$ limit generates the non-analyticity of
the free energy. For a system with crossover
$Z$ is analytic, and the zeros do
not approach the real axis as V$\rightarrow \infty$.

\begin{figure}[htb]
\centerline{\includegraphics[width=10cm]{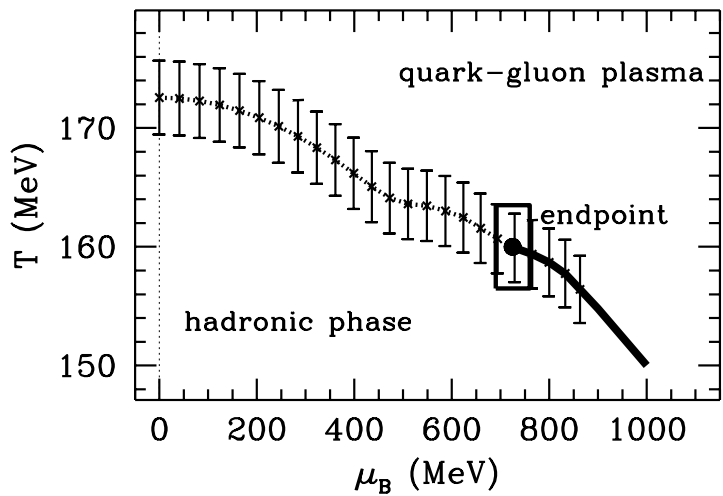}}
\caption[]{The T-$\mu$ diagram. Direct results are given with errorbars.
Dotted line shows the crossover, solid line the first order
transition. The box gives the uncertainties of the endpoint.}
\label{fig4}
\end{figure}


At T$\neq$0 we used $L_t$=4, $L_s$=4,6,8 lattices. T=0 runs were done on
$10^3\cdot$ 16 lattices. $m_{u,d}$=0.025 and $m_s$=0.2 were
our bare quark masses.
At  $T\neq 0$ we determined the complex valued Lee-Yang zeros,
$\beta_0$, for different V-s as a function of $\mu$. Their
V$\rightarrow \infty$ limit was given by a $\beta_0(V)=\beta_0^\infty+\zeta/V$
extrapolation. We used 14000, 3600 and 840 configurations on
$L_s$=4,6 and $8$ lattices, respectively.
Im($\beta_0^\infty$) is shown on Figure \ref{fig3} as a function of $\mu$.  For small
$\mu$-s the extrapolated Im($\beta_0^\infty$) is inconsistent with
a vanishing value, and predicts a crossover.
Increasing $\mu$ the value of Im($\beta_0^\infty$) decreases,
thus the transition becomes consistent with a first order phase
transition (overshooting is a finite V effect).
Our primary result is $\mu_{end}=0.375(20)$.

To set the physical scale we used a
weighted average of $R_0$,  $m_\rho$  and
$\sqrt{\sigma}$.
Note, that (including systematics due to
finite V) we have
$(R_0\cdot m_\pi)=0.73(6)$, which is at least twice, $m_{u,d}$ is
at least four times
as large as the physical values.

Figure \ref{fig4} shows the phase diagram in
physical units, thus
$T$ as a function of $\mu_B$, the baryonic chemical potential
(which is three times larger then the quark chemical potential).
The endpoint
is at $T_E=160 \pm 3.5$~MeV, $\mu_E=725 \pm 35$~MeV.
At $\mu_B$=0 we obtained $T_c=172 \pm 3$~MeV.

\section{Conclusion}

We proposed a method -- an overlap improving multi-parameter reweighting
technique -- to numerically study non-zero $\mu$ and determine the
phase diagram in the $T$-$\mu$ plane.
Our method is applicable to any number of Wilson or staggered quarks.
As a direct test we showed that for Im($\mu$)$\neq$0 the predictions
of our method are
in complete agreement with the direct simulations, whereas the Glasgow
method suffers from the well-known overlap problem.
We studied the $\mu$-$T$ phase diagram of QCD with
dynamical $n_f$=2+1 quarks.
Using our method we obtained
$T_E$=160$\pm$3.5~MeV and $\mu_E$=725$\pm$35~MeV for the endpoint.
Though $\mu_E$ is too
large to be studied at RHIC or LHC, the endpoint would
probably move closer to the $\mu$=0 axis
when the quark masses get reduced.
At $\mu$=0 we obtained $T_c$=172$\pm$3~MeV.
More work is needed to get
the final values by extrapolating
in the R-algorithm and to the thermodynamic, chiral and continuum limits.
The details of the presented results can be found in \cite{Fodor:2001au}.

This work was partially supported by 
grants 
 OTKA-\-T37615/\-T34980/\-T29803/\-M37071/\-OMFB1548/\-OMMU-708
and in part based
on the MILC collaboration's  lattice code:
http://physics.indiana.edu/\~{ }sg/milc.html.

\end{document}